\documentclass[english,journal,12pt,onecolumn,draftclsnofoot]{IEEEtran}
\usepackage[T1]{fontenc}
\usepackage[utf8]{inputenc}
\usepackage{amsmath}
\usepackage{amssymb}
\usepackage{graphicx}

\makeatletter

\newcommand{\lyxdot}{.}

\usepackage{babel}

\usepackage{babel}

\usepackage{babel}

\makeatother

\usepackage{babel}
\begin{document}
\title{Modified Zero Forcing Decoder for Ill-conditioned Channels}
\author{Ibrahim Al-Nahhal$^{(1)}$, Masoud Alghoniemy$^{(2)}$, Adel B. Abd
El-Rahman$^{(3)}$, and Zen Kawasaki$^{(4)}$\\
$^{(1)},^{(3)}$Department of Electronics and Communications Engineering,
Egypt-Japan University of Science and Technology,\\
Alexandria, 21934 Egypt (e-mail: \{Ibrahim.al-nahhal\},\{adel.bedair\}@ejust.edu.eg).\\
$^{(2)}$Department of Electrical Engineering, University of Alexandria,
21544 Egypt (e-mail: alghoniemy@alexu.edu.eg).\\
$^{(4)}$Department of Information and Communication Technology, Graduate
School of Engineering, Osaka University, Suita, \\
Osaka, Japan (e-mail: zen@comm.eng.osaka-u.ac.jp).}
\maketitle
\begin{abstract}
A modified zero forcing decoder (MZF) for Multi-Input Multi-Output
(MIMO) systems in case of ill-conditioned channels is proposed. The
proposed decoder provides significant performance improvement compared
to the traditional zero forcing for ill-conditioned channels. The
proposed decoder is the result of reformulating the $QR$ decomposition
of the channel matrix by neglecting the elements which represent the
correlation in the channel. By combining the traditional ZF with the
MZF decoders, a hybrid decoder can be formed that alternates between
the traditional ZF and the proposed MZF based on the channel condition.
We will illustrate through simulations the significant improvement
in performance with little change in complexity over the traditional
implementation of the zero forcing decoders. 
\end{abstract}


\section{Introduction}

In multi-input multi-output (MIMO) systems with additive Gaussian
noise, the Maximum-likelihood (ML) decoder is the optimum receiver.
However, due to the high complexity of the ML decoders, lattice-based
decoding algorithms such as the sphere decoding algorithms have been
proposed \cite{key-1}. Although the sphere decoder (SD) achieves
near-ML performance, its complexity is highly dependent on the received
signal to noise ratio (SNR). In order to achieve ML performance at
low SNR, the SD algorithm requires severe pre-processing operations
that involve iterative column vector ordering \cite{key-2}. On the
other hand, the zero-forcing decoder has a very low complexity at
the expense of poor performance due to its inherent noise enhancement.

In this paper, we propose a modified zero forcing decoder which provides
significant performance improvement by limiting the inherent noise
enhancement. The proposed algorithm performs a simple reformulation
of the $QR$ factorization of the channel by considering the diagonal
elements of the upper-triangular matrix $R$, which represent the
well-condition part of the channel.

Consider the complex-valued baseband MIMO model in Rayleigh fading
channels with $M$ transmit and $N$ received antennas. Let the $N\times1$
received signal vector $\dot{y}$ \cite{key-2}, 
\begin{equation}
\dot{y}=\dot{H}\dot{x}+\dot{w}
\end{equation}
 with the transmitted signal vector $\dot{x}\in Z_{C}^{M}$ whose
elements are drawn from q-QAM constellation and $Z_{C}$ is the set
of complex integers, the $N\times M$ channel matrix $\dot{H}$ whose
elements $h_{ij}$ represent the Rayleigh complex fading gain from
transmitter $j$ to receiver $i$ with $h_{ij}\sim CN(0,1)$ . In
this paper, it is assumed that channel realization is known to the
receiver through preamble and/or pilot signals, and $N\ge M$. The
$N\times1$ complex noise vector $\dot{w}$ has independent complex
Gaussian elements with variance $\sigma^{2}$ per dimension. Throughout
the paper, we will consider the real model of $(1)$ 
\begin{equation}
y=Hx+w\label{rmodel}
\end{equation}
 where, $m=2M$, $n=2N$, then $y=[R(\dot{y})\,\,I(\dot{y})]^{T}$
$\in R^{n}$, $x=[R(\dot{x})\,\,I(\dot{x})]^{T}\in Z^{m}$, $w=[R(\dot{w})\,\,I(\dot{w})]^{T}\in R^{n}$,
and $H=\left(\begin{array}{cc}
R(\dot{H}) & -I(\dot{H})\\
I(\dot{H}) & R(\dot{H})
\end{array}\right)\in R^{n\times m}$, where $R(.)$ and $I(.)$ are the real and imaginary parts, respectively.
The ML solution, $\hat{X}_{ML}$, that minimizes the 2-norm of the
residual error is found by solving the following integer-least squares
\begin{equation}
\hat{X}_{ML}=arg\,\underset{x\subset\Lambda}{min}\left\Vert y-Hx\right\Vert ^{2}\label{ML}
\end{equation}

where $\Lambda$ is the lattice whose points represent all possible
codewords at the transmitter and $Z^{m}$ is the set of integers of
dimension $m$. It should be noted that (\ref{ML}) is, in general,
NP-hard \cite{key-4}.

\section{Traditional Decoders}

In this section, we provide a brief overview of the sphere decoder
(SD) and the zero forcing decoders.

\subsection{Sphere Decoder}

The sphere decoder (SD) reduces the search complexity by limiting
the search space inside a hyper-sphere of radius $\rho$ centered
at the received vector $y$ \cite{key-4,key-5}. In particular, the
solution should satisfy 
\begin{equation}
\left\Vert y-Hx\right\Vert ^{2}<\rho^{2}\label{radius}
\end{equation}

The sphere decoder transforms the closest-point search problem into
a tree-search problem by factorizing the channel matrix $H=QR$, where
$Q$ is a $n\times m$ unitary matrix which represents the orthonormal
bases of the channel $H$, and $R$ is an upper triangular matrix
of size $m\times m$ that represents the correlation in the channel.
The energy of the residual error be written recursively as \cite{key-4}

\begin{equation}
\left\Vert y-Hx\right\Vert ^{2}=\left\Vert y-QRx\right\Vert ^{2}=\left\Vert Q^{*}y-Rx\right\Vert ^{2}<\rho^{2}\label{SD}
\end{equation}

The SD traverses the tree and computes the path metric for each node
in the tree. Any branch that has a path metric exceeding the pre-defined
pruning constraint $\rho$ will be discarded. Thus, only a subset
of the tree is visited and the complexity is reduced.

\subsection{Zero Forcing }

The zero forcing (ZF) decoder provides low complexity through decorrelating
the channel by directly inverting the channel matrix $H$ \cite{key-4,key-9}.
In particular, for non square MIMO systems, multiplying both sides
of (2) by the pseudo-inverse of the channel, the residual error can
be written as 
\begin{equation}
w_{ZF}=H^{+}w.
\end{equation}
 Where $H^{+}=(H^{*}H)^{-1}H^{*}$ is the pseudo-inverse of the channel.
It is clear that if the channel is ill-conditioned, then noise enhancement
is significant. The ZF solution, $\hat{X}_{ZF}$, minimizes the 2-norm
of the residual error. In particular \cite{key-4}, 
\begin{equation}
\hat{X}_{ZF}=\lfloor H^{+}y\rceil
\end{equation}
 where $\lfloor.\rceil$ is a slicing operation.

\section{The Modified Zero Forcing}

In order to reduce the noise enhancement which is a result of ill-conditioned
channels, the MZF only considers the diagonal elements of the upper-triangular
matrix $R$, in the $QR$ factorization of the channel matrix, $H=QR$.
It should be noted that since $Q$ is an orthogonal matrix, then $cond(H)=cond(R)$
\cite{key-12}.

In particular, let $r_{ij}$ be the $ij^{th}$ element of $R$, $j\ge i$;
and let $R=\hat{R}R_{D}$ where $\hat{R}$ is a unit upper triangular
matrix with elements $\hat{r}_{ij}={r_{ij}}/{r_{jj}},\,\,j>i$, $\hat{r}_{ii}=1$
and $R_{D}$ is a diagonal matrix whose diagonal elements are $r_{ii}$.

It should be noted that the strict upper-diagonal elements $\hat{r}_{ij}$
represent the correlation elements which are responsible for channel
bad conditioning. On the other hand, the diagonal elements $r_{ii}$
represent the energy of the channel. In order to see this, figure
$1$ illustrates the effect of decomposing the channel matrix $H$
on the condition number. In particular, the horizontal access represents
the condition number of the full channel matrix $H$, and the horizontal
access represents the condition umber of the factored matrices, $\hat{R},R_{D}$.
The blue curves represent the condition number of $R_{D}$ while the
red curves represent the condition number of $\hat{R}$ for channel
matrices of size $2\times2$ and $4\times4$. It is clear that even
for highly ill-conditioned channel matrix, the condition number of
$R_{D}$ remains low. It should be noted that figure 1 is generated
by averaging $10,000$ runs. Using the above results, the MZF decoder
solves 
\begin{equation}
\hat{X}_{MZF}=\lfloor R_{D}^{-1}Q^{H}y\rceil
\end{equation}

\begin{figure}
\centering{}\includegraphics[scale=0.65]{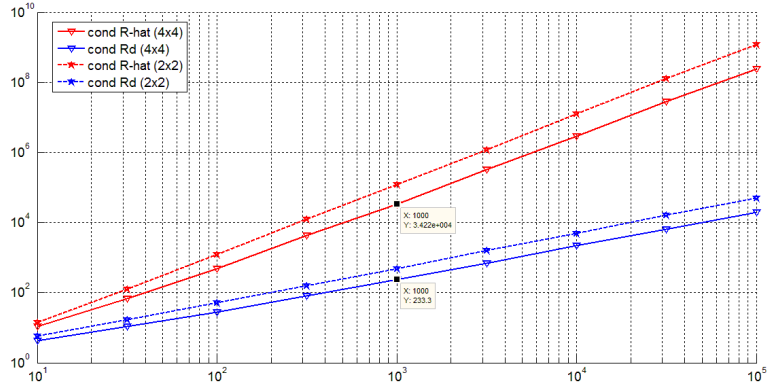}\caption{The effect of decomposition of the channel matrix H on the condition
number.}
\end{figure}

\section{Hybrid Decoder}

Based on the previous findings, a hybrid decoder (HD) can be formed
by alternating between the MZF and the ZF decoders based on channel
condition number. In particular, the MZF decoder provides performance
improvement in case of ill-conditioned channels due to considering
only the well conditioned elements of the channel. On the other hand,
in case of well-conditioned channels, the ZF decoder can be used without
loss of performance. Hence, the hybrid decoder alternates between
the traditional ZF decoder and the MZF according to the state of the
channel and can be described by the following pseudo-code.
\begin{enumerate}
\item Compute the condition number of the channel $cond(H)$. 
\item Set a threshold $\gamma$ > 1. 
\item If $cond(H)<\gamma$, then $\hat{X}_{ZF}=\lfloor H^{+}y\rceil$. 
\item If $cond(H)>\gamma$, then $\hat{X}_{MZF}=\lfloor R_{D}^{-1}Q^{H}y\rceil$. 
\end{enumerate}

\section{Complexity Analysis}

The complexity of the proposed algorithm can be measured by computing
the number of floating point operations (flops) consumed for execution;
which also can be converted into the execution time. It should be
noted that $(7)$ is solved using the $QR$ factorization algorithm
\cite{key-13}. So $(7)$ can be re-written as 
\begin{equation}
R\hat{X}_{ZF}=Q^{H}y
\end{equation}
 Which can be solved using back substitution method with cost \cite{key-13}:
\begin{equation}
flps_{ZF}\approx2nm^{2}-\frac{2}{3}m^{3}
\end{equation}
 Similarly, for the MZF decoder, 
\begin{equation}
R_{D}\hat{X}_{MZF}=Q^{H}y
\end{equation}
 The cost of the previous algorithm is the same as the cost of the
algorithm of eq. $(10)$ without the cost of back substitution. Where
the cost of back substitution is $m^{2}$, the number of flops of
MZF algorithm is: 
\begin{equation}
flps_{MZF}\approx flps_{ZF}-m^{2}
\end{equation}

In HD algorithm; the number of flops is greater than the average between
$flps_{ZF}$ and $flps_{MZF}$ by the number of flops consumed in
channel condition number calculations.

\section{Simulation Results}

For comparison purpose, the performance of the proposed decoders ,(MZF)
and (HD), are compared to the SD and the ZF for uncoded systems. It
is assumed that the transmitted power is independent of the number
of transmit antennas, $M$, and equals to the average symbols energy.
We have assumed that the channel is Rayleigh fading channel. Let $P$
be the percentage of ill-conditioned channels in the runs. Figures
2, 3, and 4 illustrate the performance of the traditional and proposed
decoders for $N=M=2$ ,$16-QAM$ for well-conditioned channels $P=0\%$,
ill-conditioned channels with $cond(H)=10^{3}$ for $P=100\%$ and
$P=50\%$ versus different values of signal to noise ratio. Figure
5 illustrates the performance for different values of channel condition
numbers for $SNR=15dB$. As it is clear from the figures, the SD is
superior while the ZF and the MZF decoders are inferior. It should
be noted that in the well-conditioned channel case, as illustrated
in figure 2, the MZF decoder performance has an error floor. This
is a natural byproduct of neglecting of one of the well-conditioned
channel component, $\hat{R}$, as expected. Thus, the HD typically
follows the ZF because it acts as a ZF in well-conditioned channels,
$P=0\%$, as indicated before. Similarly, for the ill-conditioned
channels, $P=100$, illustrated in figure 3, the performance of the
ZF decoder produces an error floor. This is expected due to the noise
enhancement inherent in the ZF decoder \cite{key-12}. Figure 4 shows
that as the percentage of ill-conditioned channels increases, the
performance of HD is very close to SD performance especially in low
SNR. Also, as shown in figures 5 and 6 the performance of the HD approaches
the SD performance with the increase of the channel condition number
especially in low SNR as shown in figure 6 .

The complexities of the SD, ZF, MZF, and HD are measured by the execution
time in finding the solution. In particular, figures 7 illustrates
the complexity for the $2\times2$ MIMO with $16-QAM$ modulation
as a function of the SNR. It is clear that the SD has high complexity
while the MZF and the ZF decoders has low complexity. As it is clear
from $(12)$, the complexity of MZF decoder is less than the complexity
of the ZF decoder by a small margin according.

\begin{figure}
\centering{}\includegraphics[scale=0.65]{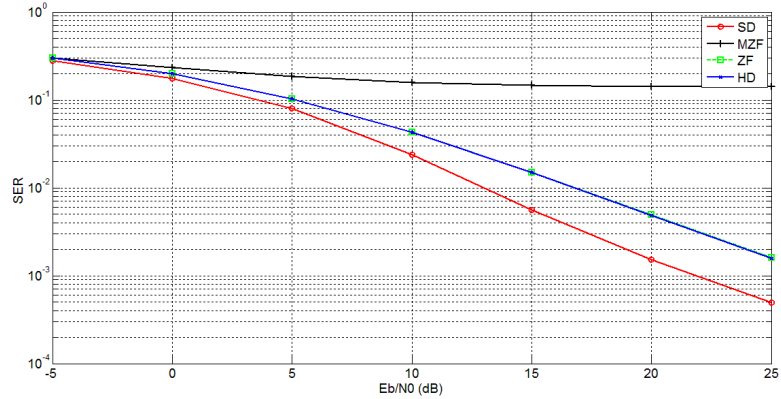}\caption{Performance of 2$\times$2 MIMO 16-QAM well-conditioned channels (P
= 0\%)}
\end{figure}

\begin{figure}
\centering{}\includegraphics[scale=0.65]{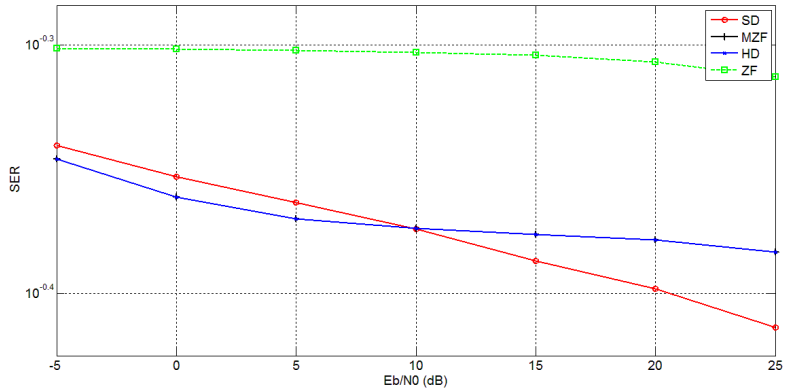}\caption{Performance of 2$\times$2 MIMO 16-QAM ill-conditioned channels with
cond(H) = $10^{3}$ (P = 100\%)}
\end{figure}

\begin{figure}
\centering{}\includegraphics[scale=0.65]{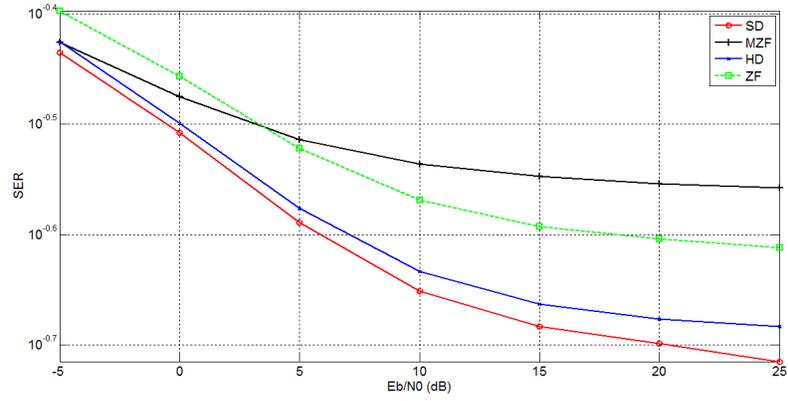}\caption{Performance of 2$\times$2 MIMO 16-QAM (P = 50\%)}
\end{figure}

\begin{figure}
\centering{}\includegraphics[scale=0.65]{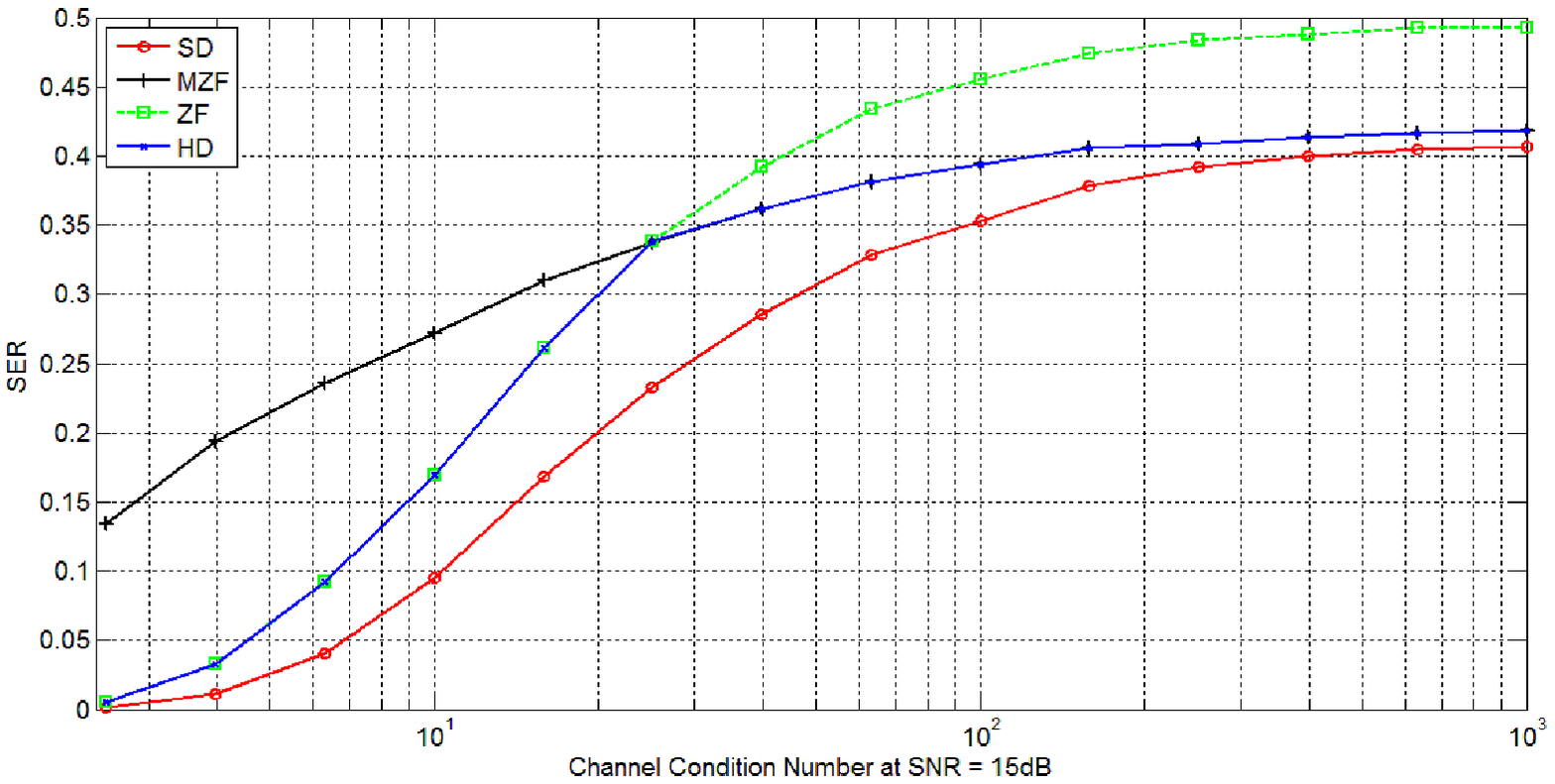}\caption{Channel performance varies with a channel condition number for 2$\times$2
MIMO 16-QAM at SNR = 15dB}
\end{figure}

\begin{figure}
\centering{}\includegraphics[scale=0.65]{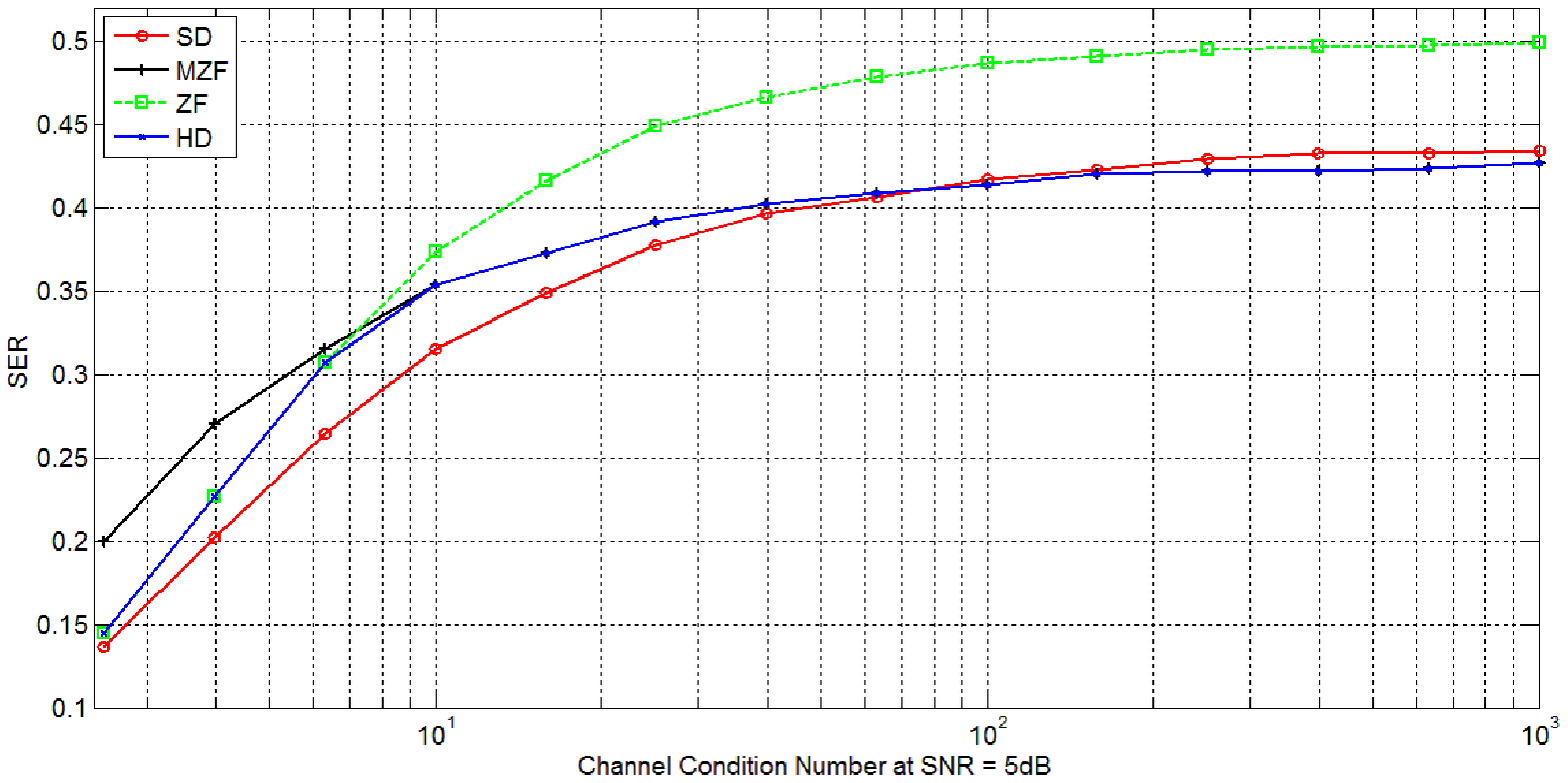}\caption{Channel performance varies with a channel condition number for 2$\times$2
MIMO 16-QAM at SNR = 5dB}
\end{figure}

\begin{figure}
\centering{}\includegraphics[scale=0.65]{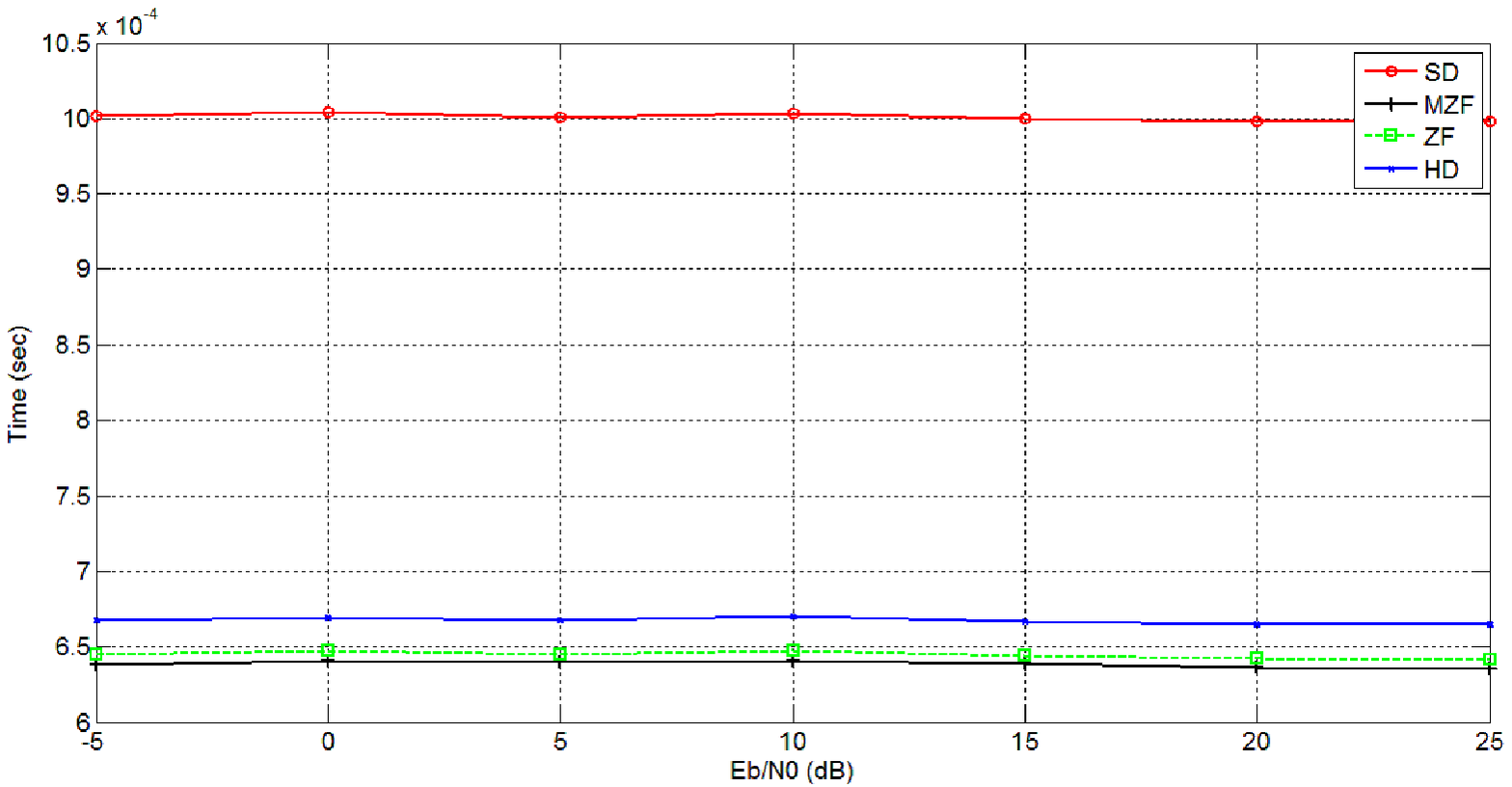}\caption{Average Execution Time per Bit for 2$\times$2 MIMO 16-QAM}
\end{figure}

\section{Conclusions}

A Hybrid MIMO decoder that is based on neglecting the cross correlation
elements of the channel correlation matrix, $R$, in ill-conditioned
channels and acting as ZF in well-conditioned channels has been proposed.
The proposed decoder has better performance in ill-conditioned channels
than the corresponding ZF decoder. The complexity of the proposed
decoder is as low as the ZF decoder.

\end{document}